\documentclass[prd,showpacs,letterpaper,10pt,aps,notitlepage,nofootinbib,superscriptaddress]{revtex4-2}
\usepackage[utf8]{inputenc}
\usepackage{color}

\usepackage{physics}
\usepackage{slashed}
\usepackage{amsmath}
\usepackage{mathtools}
\usepackage{placeins}
\usepackage{amsfonts}
\usepackage{amssymb}

\usepackage{hyperref}
\hypersetup{colorlinks, linkcolor = [rgb]{0,0.0,0.75}, citecolor = [rgb]{0,0.0,0.75}, urlcolor = [rgb]{0,0.0,0.75}}

\begin{document}

\title{Theory of CP angles measurement\footnote{ Invited talk at the KM50 and Flavor Factory 2023 workshop held at KEK Feb 9-11, 2023 to celebrate  50th anniversary of the  paper (1973)  by M. Kobayashi and T. Maskawa on Theory of CP violation and Flavor Factory 2023}}



\author{Amarjit Soni}
\affiliation{Physics Department, Brookhaven National Laboratory, Upton, NY 11973, USA}


\begin{abstract}
In the early 80's Sanda-san and collaborators wrote  key papers on  the direct and clean determination  of the unitarity angle $\phi_1$ ($\beta$). This motivated many of us for analogously coming up with  ways for direct and clean determinations of the other two unitarity angles, $\phi_2 (\alpha)$ and $\phi_3 
(\gamma)$.                                     
Current status of these direct determinations as well as our expectations for when 
Belle-II has 50 $ab^{-1}$ of luminosity and LHCb with some upgrades, will be given. 
In particular, it is emphasized that for direct determination of $\phi_3$,  Belle-II should be able to handle final states in $D^0$ or $\bar D^0$ Dalitz decays, that  contain one $\pi^0$ (which are difficult for LHCb) then they may make further inroads in improving the accuracy of $\phi_3$ determination.
Early lattice inputs for constraining the unitarity triangle (UT) are briefly recalled. Its crucial role in supporting the Kobayashi-Maskawa theory of CP violation is emphasized.
Over the years lattice methods have made significant progress and latest constraints from these for the UT will be discussed as well as compatibility with current direct determinations and some comments on future outlook will be made.
\end{abstract}

\maketitle

\section{Introduction}

We are here to celebrate 50 years of the remarkable paper~\cite{KM72} by Kobayashi and Maskawa, a paper that for many of us has been the focus for the past many decades. A lot of impetus for clean and direct determinations of angles of the unitarity triangle (UT) came from the celebrated paper by Bigi and Sanda~\cite{Bigi:1981qs} in which they focused on $\phi_1 (\beta)$ through the process $B^0 (\bar B^0 ) \to
\psi K_s$. That work was very important for motivating the construction of
B-factories. But a key role in making an asymmetric B-factories was also played by Pier Oddone [See the proceedings of a B Physics workshop held at UCLA around 1985] who came up with the idea of the asymmetric B-factory to make use of the long life time of the B meson. We'll review here the clean and direct determinations of all three angles of the UT and where they stand as of now.

It is also useful to briefly discuss here the indirect determinations of the UT.
Around 2001 when the B-factories were just starting it was shown~\cite{Atwood:2001jr} for the first time how four observables from the lattice suffice to provide a constraint on $\sin(2\phi_1)$ yielding  roughly compatible value to what was being determined from experiment within largish errors. On the lattice, significant progress was made in the next $\approx$ 5 years and by 2007 it was clearly demonstrated, in conjunction  with indirect CPV seen in K decays and O(1) CPV seen in B-decays that both can  be accounted for by the KM theory. For this huge accomplishment, Kobayashi and Maskawa were awarded the Nobel Prize in 2008. In this context it is important to remember that there is no other way to reliably say that the indirect CP violation in K-decays comes from the same phase without precise input from the lattice for the kaon B-parameter, $B_K$.

\section{Direct determination of the angles of the Unitarity Triangle (UT)}          

This topic got initiated by the seminal papers of Sanda-san~\cite{Carter:1980hr,Carter:1980tk,Bigi:1981qs}. While strictly speaking these papers focus on direct determinations of only $\phi_1 (\equiv \beta$), they motivated many of us to analogously search for ways for direct determinations of the other two angles $\phi_2 (\equiv \alpha$) and $\phi_3 (\equiv \gamma$). 

The key observation of Bigi and Sanda~\cite{Bigi:1981qs}  was that the process $B \to \psi K_s$ receives contributions from two paths:
First there is the decay, $B or \bar B \to \psi Ks$. In the Wolfenstein representation\cite{Wolfenstein:1983yz},  in the leading order 
in $\lambda \equiv \theta_{Cabibbo}$ this weak decay has no CP-phase.
But then there is also the second path. In this case $B^0$ oscillates  to $\bar B^0$ and then decays to $\psi Ks$. The oscillation of $B^0$ to $\bar B^0$ does involve a substantial CP phase i.e. $\sin 2 \phi_1$,  moreover, as emphasized in Sanda-san's papers, this is a remarkably clean way of directly accessing from experiment the underlying CP-phase. 

After many years of experimental and theoretical study of time dependent CP violation in the mode, $B \to \psi K_s$, the current world average stands at (See the very nice review by T. Gershon et al on "Determination of CKM angles from B hadrons" in~\cite{ParticleDataGroup:2022pth})

\begin{align}
    -\eta_{CP} S_f = 0.699 \pm 0.017    ; ~ ~ ~ ~ ~   C_f = -0.005 \pm 0.015
\end{align}

\noindent where $S_f$ is the coefficient that  monitors the time-dependent CP-violation whereas $C_f$ is the  coefficient of the time independent i.e. direct CP violation term.

It should be stressed, as is done in the review by Gershon et al~\cite{ParticleDataGroup:2022pth} that  despite the years of data taking by BABAR, Belle and LHCb, the errors in the above determinations are completely dominated by statistics. Higher order corrections,  some times also called penguin pollution, have been estimated~\cite{Ciuchini:2005mg,Faller:2008zc,Ligeti:2015yma} and found to be small in comparison to the current
statistical errors.  Presumably, Belle-II and LHCb upgrades will make advances here and it is likely that 
in the coming decade systematic errors will receive greater scrutiny.

Now let us briefly discuss the direct determination of the other two angles: $\alpha \equiv \phi_2$
and $\gamma \equiv \phi_3$.

For direct extraction of $\phi_2$ there are several methods.\\ 

First among these is the Gronau and London use of the "isospin triangle" for $B \to \pi \pi$.  ~\cite{Gronau:1990ka} .  Then there  is the $\rho \rho$ method.  This has some advantages over the two $\pi$  method, especially since the  $\rho\rho$ Br is a lot bigger.
However,  these methods are hampered  by  discrete ambiguities.   Time dependent CP violation  by 
the $\rho\pi$ method leading to three pions  in the  final state,  has the advantage that there are five amplitudes and one can use   the power of the Dalitz plot and the method does not have  discrete ambiguities~\cite{Snyder:1993mx}. But it seriously suffers, till this day from  lack of statistics. This will therefore be one arena where Belle-II  and the LHCb upgrades can  have important impact. 

Be that as it may as shown clearly  in the PDG 2022 review article by Gershon et al~\cite{ParticleDataGroup:2022pth}, see in particular Fig. 77.2  the central value  World Average of
$\alpha \equiv \phi_2 = 85.2$
degrees. The current error of about 5.5\% is dominated by statistics. The systematics
are estimated only about 1\%. The effects of EWP, isospin breaking and of $\rho$
finite width are being neglected.  In the coming years with much larger data sets expected from BELLE-II and from LHCb upgrades significantly better determination of 
$\phi_2 \equiv \alpha$ should take place. \\

      For $\phi_3 (\gamma)$, let us first briefly recall Gronau and Wyler~\cite {Gronau:1991dp} seminal
      idea that $b\to c$ and $b \to u$ interference leading to CP-eignenstates of $D^0$ and $\bar
      D^0$ such as $K_S \pi^0$, $K^+ K^-$, $\pi^+\pi^-$ etc via $B^- \to K^- D^0$ or
      $K- \bar D^0$. This is clearly an excellent idea that can be exploited to deduce directly from the data the unitarity angle $\phi_3$. But given that one of these is color allowed
      ($B^- \to K^- D^0$) and the other color suppressed there is a concern that the interference and therefore the size of CP asymmetry may be limited.

In their attempt to  enhance the interference and thereby the CP asymmetry 
ADS~\cite{Atwood:1996ci} focused on CP non-eigenstates of $D^0$. To this end, the color-allowed decay $B^- \to K^- D^0$ is followed by doubly-Cabibbo suppressed decay (as an example), $D^0 \to K^+ \pi^-$ whereas the color suppressed mode $B^- \to K^- \bar D^0$ is followed by the Cabibbo allowed 
decay, $\bar D^0 \to K^+ \pi^-$. By using several different decays of $D^0$
to $K^- \pi^+$, $K^- \rho^+$, $K^{*-}\pi^+$ etc. as explicitly shown in~\cite{Atwood:1996ci}
(see in particular eq. 3 therein) the angle $\phi_3$ that we are after can be obtained by solving
algebraic equations (along with the unknown strong (CP-even phases))  directly from the data and the accuracy of this determination is completely data driven.  

Note that in these $\gamma$ determinations, charged $B^+$ and $B^-$ decays are being used
 (actually $B^0$ and $B_s$
can also be used), so one is dealing with manifestly direct CP violation and one is able to use hadronic decay products to deduce directly from the data simultaneously both strong (CP-even) and weak (CP-odd) phases.
In direct CP violation for K-decays this is not possible and this is why  phenomenological methods and lattice approaches have to be used~\cite{Buras:2022cyc, Aebischer:2021hws,Cirigliano:2019obu, 
Cirigliano:2019zjv, Bertolini:1998vd,RBC:2020kdj,Blum:2023mtn}.

As  already suggested briefly in the ADS-PRL~\cite{Atwood:1996ci}, for $D^0,(\bar D^0)$ into 3-body modes Dalitz plot analysis can be very powerful. Both GGSZ~\cite{Giri:2003ty} and ADS~\cite{Atwood:2000ck} point different approaches of implementing model independent Dalitz analysis.
Since GGSZ was already discussed by Matthew Kenzie, couple of points about ADS (see esp. Section  Vi
in~\cite{Atwood:2000ck}) are covered here.

ADS first looked for minimum values of $\phi_3$ in local regions of the Dalitz  plot and then searched for a "global" minimum among them.

In the second step ADS implemented the method of "optimum observable"~\cite{Atwood:1991ka} and showed that their "global" minimum value for $\gamma$ was close to that obtained by the optimum
observable method.

Briefly when extracting the value of any parameter from the data given many possible observables,
there is an optimal observable~\cite{Atwood:1991ka} whose explicit construction is given in that paper,  which is rigorously proven  to have the minimum statistical error.  
As is well known, in recent years Machine Learning
techniques have become very popular in Particle Physics.  
At the core of the algorithms that are used in that context is in fact this theorem 
on “optimal observable” proved in this paper in 1992; see, for example,~\cite{Brehmer:2019xox} in particular their reference 27.

In recent years considerable focus  has been on the GGSZ~\cite{Giri:2003ty} mode $D^0 \to K_S \pi^+ \pi^-$, which gets fed by $K^+\rho^-$, $K_S \rho^0$, $K^{*+}\pi^-$, a mode  that was also actually mentioned briefly in~\cite {Atwood:2000ck}. However, following the then experimental data that was available from Fermilab Experiment E687~\cite{E687:1994wlh}, ADS presented a detailed Dalitz analysis of the mode $D^0 \to K^-\pi^+\pi^0$ which can, of course, be fed by $K^{*-} \pi^+$, $K^{*0}\pi^0$, $K^- \rho^+$ etc. The presence of a $\pi^0$ certainly makes the analysis somewhat more complicated. But the final state has only one $\pi^0$. It is likely that this is less of a problem for Belle-II, so they may use this opportunity to make further inroads into improving the accuracy of $\phi_3$ determination. 

Fig 77.3 in
[See again the review by Gershon et al on "Determination of CKM angles from B hadrons"] ~\cite{ParticleDataGroup:2022pth} shows CL for $\gamma$ from many different modes. The 
World average of $\phi_3$ = $\gamma = 65.9^{\circ}$  with an error of   about $3.5^{\circ}$
Again, the statistical error dominates. In particular, Brod and Zupan~\cite{ Brod:2013sga} have emphasized that
higher order corrections to $\gamma$ are very tiny so pushing  for a reduction in statistical errors is well motivated and likely to stay that way through the period of BELLE-II reaching its  luminosity goal of 50/ab and and so also with LHCb upgrades.  


\begin{table}[htb]
\caption{The current status and future outlook for the angles of the UT via  {\it direct} determinations.
Numbers given are in degrees.
Assumptions being made for future extrapolations are further explained in text.}

 \label{tab:direct}
\begin{tabular}{c|c|c|c}
  angle & current &  future  (Lumi $\approx50 ab^{-1})$ & comments. 
\tabularnewline
\hline \hline
$\phi_1 (\beta)$ & $22.2 \pm 0.7 $ & $\pm 0.07$ & assumed a factor of 100 in eff. Lumi  \tabularnewline
\hline 
\
$\phi_2 (\alpha)$ & $85.2^{+4.8}_{-4.3}$ & $\pm 0.3$ & assumed a factor of 225 in  eff. Lumi \tabularnewline
\hline 
\
$\phi_3 (\gamma) $ & $65.9^{+3.3}_{-3.5}$ &  $\pm 0.17$ & assumed a factor of 400  in eff. Lumi \tabularnewline
\hline \hline
\end{tabular}
\end{table}



Table~\ref{tab:direct} summarizes where we are now with the direct methods and indicates
the accuracy in future ($\approx$ 10 years time). In these extrapolations it is assumed that 
Belle-II integrated luminosity will gain by a factor of fifty and quite possibly similar
increase will result from LHCb upgrades. Thus, for $\phi_1$ a reduction in statistical error by
a factor of ten is anticipated. For $\phi_2$ a factor of 15 in reduction of statistical error 
is assumed bearing in mind in part the power of the Dalitz method. For $\phi_3$ not only Dalitz plots but also multiple channels should all possibly give a factor of $\approx$ 20 reduction in statistical error.

\section{Indirect determination of angles of the Unitarity Triangle}
\label{Indirect determinations}

For indirect determination of UT,  experimental data plus  input from the lattice and from phenomenology  are all used. Important observables were identified and their extractions from 
the lattice in exploratory calculations was dealt in these rather early calculations~\cite{Bernard:1994kp,Bernard:1993zh,Bernard:1991bz,Bernard:1991fg}.
Use of SU(3) breaking ratio, $\xi$, which can be determined from the lattice. in conjunction
with the experimental measurements of $B_s$ oscillation frequency, when that becomes available
(see below) was proposed in~\cite{Bernard:1998dg,Kronfeld:2002ab}.

The overall strategy for constraining UT angles from the lattice input
was spelled out in ~\cite{Atwood:2001jr} and presented at the “B physics and CP  violation”, meeting organized by Sanda-san for Feb 2001 held at Ise (Japan). Crucial  entities that play the key role were identified. These were:

\begin{itemize}
\item  info from semi-leptonic decays for $|V_{ub}/V_{cb}|$
\item $f_B \times \sqrt (\hat B_B)$
\item $\hat B_K$
\end{itemize}
Table~I in~\cite{Atwood:2001jr}  summarizes the early lattice results based mostly on naive Wilson fermions which do not respect chiral symmetry of the continuum theory especially on the coarse lattice  spacings that 
 were then available.. This is why the $B_K$ error
is rather large (by today's standards (see later)). Nevertheless, it is remarkable that these results available just around the time the two asymmetric B-factories started to take data showed that $\sin 2 \phi_1$  within largish errors is quite compatible with early direct measurements (also with appreciable errors).

In the next ~5 years B-factories made some progress in direct determinations but their
progress  in providing
vital experimental information, (for example, in semi-leptonic decays and in $B_d$ mixings) was even more significant. 

Around 2005-06 for the B-physics UT a very important measurement came from a hadron collider. CDF gave a remarkably   precise determination of the extremely challenging $B_S$ oscillation frequency:

\begin{align}
\Delta m_s = 17.31^{+0.33}_{-0.18} (stat.) +- 0.07 (syst.) ps^-1
\end{align}
Using the masses of $B_d$ and $Bs$ mesons and the  corresponding mass differences $\Delta m_d$ and $\Delta m_s$ and from the lattice the SU(3) breaking ratio $\xi$`\cite{Bernard:1998dg,Kronfeld:2002ab} leads to a rather accurate
determination of the ratio of mixing angles

\begin{align}
|V_{td}/V_{ts}| = \xi \sqrt((\Delta m_d/\Delta m_s)(m_{B_s}/m_{B^0}))
\end{align}

On the lattice side while there was steady progress across the board, 
on $B_K$
pivotal developments started 1996-97~\cite{Blum:1996jf,Blum:1997mz} with the use of domain-wall quarks which preserve chiral symmetry to a high degree of accuracy.
Finally in 2005-2007 these led to a determination of $B_K$ in full (i.e. not-quenched ) QCD with the total error $\le 10\%$~\cite{RBC:2007bso}. This was crucial in demonstrating unambiguously for the first time that the CP 
asymmetry seen in $K_{L} \to 2 \pi$ of about $10^{-3}$ and the O(1) CP asymmetry seen in B-decays,
for example, in $B \to \psi K_s$ were simultaneously accounted for by the single phase
$\eta$ (in the Wolfenstein~\cite{Wolfenstein:1983yz} representation) by the KM theory of CP violation that culminated in 2008 Nobel 
Prize for Kobayashi and Maskawa.

It is useful to recall that $B_K$ defined as~\cite{RBC:2007bso}
\begin{equation}
B_K = \bra K  ( \bar s \gamma_{\mu} (1 - \gamma_5)d)( \bar s \gamma_{\mu} (1 - \gamma_5)d)\ket {\bar K} ( 8 f_K^2 m_K^2)/3)
\end{equation}

\noindent where $f_K \approx 155 MeV$ is the pseudoscalar decay constant of the  kaon and $m_K$ is its mass, is just a theoretical construct which is inherently of a non-pertutbative nature that is needed
to compare the theoretical prediction for the indirect CP violation in $K_L \to 2 \pi$ from the SM with experiment. In the absence of the lattice,
flavor SU(3) plus leading order chiral perturbation theory, using life-time of $K^+ \to \pi^+\pi^0$ gives an estimate~\cite {Donoghue:1982cq} 
but there is no clear understanding of the relevant scale, $m_{\pi}$, $m_K$ or $\mu$,
the renormalization scale. Then there is also, of course, a prediction based on large $N$~\cite{Bardeen:1987vg}
and one based on ChPT + large N~\cite{Pich:1995qp}.

One of the major advantage of the lattice approach to evaluating such matrix elements in sharp contrast to above methods is that once with the lattice an approach becomes known, it is systematically
improvable. Since the 2007 result with DWQ, many different discretizations~\cite{Aoki:2010pe,Aubin:2009jh,BMW:2011zrh} have been used and the error on $B_K$ is now down to around 1.3\%~\cite{FlavourLatticeAveragingGroupFLAG:2021npn}.

In this regard (that with the lattice) the errors are systematically improvable can be gleaned
from Table{current}, The error on $\xi$ is now down to about 1.5\%, $|V_{ub}/V_{cb}|$ to about 5\%
and on $f_B$ to around 2.2\%. Thus in about 15-20 years
these errors are down about a factor of 4 to about 12.

        

\begin{table}[htb]
\caption{Indicates progress made by the lattice in about 20 years. 2021 numbers are taken from~\cite{FlavourLatticeAveragingGroupFLAG:2021npn}}.

\label{tab:lattice_Flag}
\begin{tabular}{c|c|c}
 quantity& error around 2001 & error around 2021 \tabularnewline
\hline \hline
$\abs {V_{ub}/V_{cb}}$ &  20\% &  5\%
\tabularnewline
$f_{B_d}\sqrt{\hat B_{B_d}}$ & 22\% & 2.2\% 
\tabularnewline
$\xi$ & 7.1\%  & 1.5\%
\tabularnewline
$\hat B_K$ & 17\% & $1.3\%$ \tabularnewline
\hline \hline
\end{tabular}
\end{table}

The upshot is that tne B-UT ("Standard" UT)  deduced by direct methods is quite compatible with the one deduced by indirect methods.
However, some issues remain unresolved for quite some time. $V{xb}$ (for $x=u,c$) determined by
inclusive (non-lattice) methods is incompatible with that determined by exclusive (i.e.) lattice methods by about 2 to 3 $\sigma$~\cite{ParticleDataGroup:2022pth}. Also tension(s) of O(2 $\sigma$) persist in the UT~\cite{Lunghi:2010gv,Cornella:2021sby},

\section{Summary}
The paper by Kobayashi-san and Maskawa-san on the theory of CP violation is truly remarkable. It accounts for all aspects of CP violation in K, D, B to the extent reliable theoretical predictions can be made. Its influence on our field may last for a long time to come. But of course we should continue to improve the precise evaluation  of the angles of the unitarity triangle, $\phi_1, \phi_2$ and $\phi_3$ with  both methods direct and indirect. This is one of the best inclusive way to look for new phenomenon. To the extent that the angles do or do not add to 180 degrees will be indicative of the existence of a new phase.

While in the direct methods especially for $\alpha$ and $\gamma$ higher fluxes of B, $B_s$
in the next 5-10 years because of Belle-II anticipated increase in luminosity as well as with LHCb upgrades, significant strides are expected. For $\phi_3$, Belle-II will have an additional opportunity  via the $D^0$ Dalitz decay modes involving one $\pi^0$ in the final state  that ADS~\cite{Atwood:2000ck} discussed in  and also mentioned in brief above when discussing direct extraction of $\phi_3$. For LHCb, presence of $\pi^0$ was problematic. This would be a very good opportunity 
for Belle-II to further improve $\phi_3$.

In so far as indirect methods go, additional experimental data sets along with much faster computers 
should likely help in further reducing the errors in lattice calculations. Progress is expected in
the current few $\sigma$ tensions between exclusive and inclusive methods for $V_{cb}$ and $V_{ub}$~\cite{Aoki:2023qpa}

\section{Acknowledgements}
The author wishes to thank Shoji Hashimoto, Takashi Kaneko and other organizers for this wonderfully organized event. Special thanks also to Tom Browder, Alan Schwartz and Oliver Witzel for many helpful converstations.

\bibliographystyle{hunsrt.bst}
\bibliography{KM50}

\end{document}